\documentclass{PoS}

\usepackage{graphics}
\usepackage{bm}
\usepackage{bbm}
\usepackage{yfonts}
\usepackage{array}
\usepackage{amssymb}
\usepackage{amsopn}
\usepackage{amsmath} 
\usepackage{mathrsfs}
\usepackage{lmodern}
\usepackage{url}

\title{Monte Carlo Techniques in small-x Physics: Formal Studies and Phenomenology}

\ShortTitle{Monte Carlo Techniques in small-x Physics}

\author{\speaker{G. Chachamis}\\
    Instituto de F\'{\i}sica Corpuscular, Universitat de Val\`encia -- 
Consejo Superior de Investigaciones Cient\'{\i}ficas,
Parc Cient\'{\i}fic, E-46980 Paterna (Valencia), Spain\\  
        E-mail: \email{grigorios.chachamis@ific.uv.es}}

\author{A. Sabio Vera\\
        Instituto de F\'isica Te\'orica UAM/CSIC, Nicol\'as Cabrera 15, UAM, E-28049 Madrid, Spain\\
        E-mail: \email{agustin.sabio@uam.es}}

\abstract{We discuss the solution to the BFKL equation in the adjoint representation 
at LO and NLO accuracy for the $\mathcal{N} = 4$ SUSY theory.
We use Monte Carlo techniques to study numerically the Gluon Green's function at LO and NLO  directly written in the transverse momentum space 
which allows for the factorization of its infrared divergencies. Finally,
we discuss the applicability of our approach
to phenomenological searches for the BKP Odderon  at the LHC.}

\FullConference{XXI International Workshop on Deep-Inelastic Scattering and Related Subject -DIS2013,\\
		22-26 April 2013\\
		Marseilles,France}

\begin{document}

\section{Introduction}

The Balitsky-Fadin-Kuraev-Lipatov (BFKL) framework at leading (LO)~\cite{BFKL1}
and next-to-leading (NLO)~\cite{Fadin:1998py} logarithmic accuracy has been used to study
the properties of scattering amplitudes in Quantum Chromodynamics (QCD) and ${\mathcal N} = 4$ supersymmetric Yang-Mills theory (MSYM) in certain kinematic regions (multi-Regge (MRK) and 
quasi-multi-Regge kinematics) where logarithms of the center-of-mass energy are enhanced.
Physical observables such as hadron structure functions at small values of Bjorken $x$ 
in deep inelastic scattering or inclusive dijet production with a significant rapidity separation at the Large Hadron Collider are characteristic cases~\cite{Vera:2006un,Bartels:2006hg,Chachamis:2009ks,Chachamis:2011rw,arXiv:1106.6172,arXiv:1110.5830,arXiv:1110.6741}
where the  BFKL approach is suitable.

From a more formal point of view, it was in a generalized leading logarithmic approximation, 
where the Bartels-Kwiecinski-Praszalowicz (BKP) equation was 
proposed~\cite{Bartels:1980pe,Kwiecinski:1980wb} and found to have a hidden 
integrability~\cite{Lipatov:1985uk,Lipatov:1990zb,Lev1,Lev2,Lipatov:1994xy,Faddeev:1994zg,Lipatov:2009nt,Bartels:2011nz,arXiv:1111.4553}. 
Moreover, corrections to the 
Bern-Dixon-Smirnov (BDS) iterative ansatz~\cite{Bern:2005iz}  were found 
in MRK and within the BFKL formalism
 in~\cite{Bartels:2008ce,Bartels:2008sc}. These corrections have been understood as part of the finite remainder to the amplitude which corresponds to the anomalous contribution of a conformal Ward identity~\cite{Drummond:2007au,Bern:2008ap,DelDuca:2010zg,Goncharov:2010jf,Lipatov:2010ad,Bartels:2010tx,Bartels:2011xy,Dixon:2011pw}.

The main bulk of the literature focuses on seeing the BFKL dynamics 
as the dynamics of the BFKL Pomeron, that is, a colorless state 
of two reggeized gluons exchanged in the $t$-channel.
Nonetheless though,  
the BFKL framework is applicable to any color state the two reggeized gluons might be in. 
For three colors, that is, for $N_c = 3$, there are six irreducible representations: 
$\underline{1}, \,\underline{8_a}, \,\underline{8_s}, \,\underline{10}, \,\underline{\overline{10}}, \,\underline{27}$.
From a purely theoretical point of view,
both the symmetric and antisymmetric color octet or adjoint representation 
are extremely important. Firstly, because of the gluon reggeization and secondly, because
the color octet representation is connected to the BDS ansatz and to the BKP equation.
It is mainly the latter that interests us in this work.
We should mention
that we use the terms `adjoint' and `color octet' representation interchangeably,
by both we mean the same thing.  Our analysis for the color
octet state can in principal be extended to any color state.

In Refs.~\cite{Chachamis:2011nz,Chachamis:2012fk} we used advanced Monte Carlo techniques~\cite{code} 
to study the exclusive information present in the gluon Green's function in 
the octet representations at LO and NLO
accuracy. Some of the results are repeated here so that we can
make the connection between the insight gained by our previous
work to an ongoing project that has in its core the phenomenological
study of the BKP Odderon at the LHC.
We present the iterative solution in Section 2 for both the LO and NLO cases.
We spare all technical details for which we refer the reader to the original
publications~\cite{Chachamis:2011nz,Chachamis:2012fk}.  We conclude and
present an outlook in Section 3.

\section{The adjoint BFKL Green function at LO and NLO in iterative form}
The non-forward BFKL equation for a general color representation 
at leading and next-to-leading order, can be found
in Ref.~\cite{Fadin:2000kx,arXiv:1111.0782,Fadin:2005zj}. 
In this section we are interested in comparing 
the gluon Green's function in the octet color state against that of the singlet color state. 
The only 
difference between both solutions is in the ``real emission'' part of the kernel, which 
in the color octet case carries an extra factor of $1/2$ with respect to the singlet case. 
In the octet case, this spoils the complete cancellation 
of infrared divergencies present in the singlet, or Pomeron, projection.  

We can show that the extra infrared divergencies that appear in the non-singlet 
representations can be written as a simple overall factor in the gluon Green's 
function. For that we regularize half of the divergencies in the gluon Regge trajectory 
using dimensional $D=4-2 \,\epsilon$ regularization, while the remaining ones are 
treated using a mass parameter $\lambda$, which is also used to regularize the 
phase space integral of the ``real emission'' sector. The dependence on $\lambda$ 
will cancel out while the dependence on $\epsilon$ will remain in the factorized term. 
We use the notation ${\bf q}_i' \equiv {\bf q}_i - {\bf q}$, where ${\bf q}$ is the 
momentum transfer and all two--dimensional vectors are represented in 
bold.

From now on we will focus on the description of the infrared finite remainder 
of the gluon Green's function. We factor out the $\epsilon$ dependence and in order 
to have the singlet and octet solutions we set 
set  $c_{\cal R} = 1$ and $c_{\cal R} = 1/2$ respectively. 
The finite remainder then at LO reads:

\begin{eqnarray}
{\cal H} \left({\bf q}_1,{\bf q}_2;{\bf q};{\rm Y}\right) &=& 
\left(
{\lambda^{2 } \over \sqrt{{\bf q}_1^2 {\bf q}_1'^2} }
\right)^{{c_{\cal R} {\bar \alpha}_s}{\rm Y}} 
\Bigg\{\delta^{(2)} \left({\bf q}_1-{\bf q}_2\right) \nonumber\\&&\hspace{-3.2cm}+\sum_{n=1}^\infty \prod_{i=1}^n c_{\cal R} \int 
{d^2 {\bf k}_i \over \pi {\bf k}_i^2} \theta ({\bf k}_i^2-\lambda^2) 
\,  {{\bar \alpha}_s \over 2} 
\Bigg(1+{\left({\bf q}_1'+\sum_{l=1}^{i-1}{\bf k}_l\right)^2 ({\bf q}_1+\sum_{l=1}^{i}{\bf k}_l)^2 - {\bf q}^2 {\bf k}_i^2 \over 
({\bf q}_1'+\sum_{l=1}^{i}{\bf k}_l)^2 \left({\bf q}_1+\sum_{l=1}^{i-1}{\bf k}_l\right)^2}\Bigg)\nonumber\\
&&\hspace{-3cm}\times
\int_0^{y_{i-1}} \hspace{-0.4cm} d y_i \left(\left({\bf q}_1+\sum_{l=1}^{i-1}{\bf k}_l\right)^2
\left({\bf q}_1'+\sum_{l=1}^{i-1}{\bf k}_l\right)^2 \over 
\left({\bf q}_1+\sum_{l=1}^{i}{\bf k}_l\right)^2
\left({\bf q}_1'+\sum_{l=1}^{i}{\bf k}_l\right)^2 \right)^{{{\bar \alpha}_s \over 2}y_i}\hspace{-0.6cm} \delta^{(2)} \left({\bf q}_1+\sum_{l=1}^n {\bf k}_l-{\bf q}_2\right) \Bigg\},
\label{H}
\end{eqnarray}
whereas at NLO in the adjoint representation it explicitly reads:
\begin{eqnarray}
{\cal F} \left({\bf q}_1,{\bf q}_2;{\bf q};{\rm Y}\right) &=& 
\left({{\bf q}^2 \lambda^{2} \over {\bf q}_1^2 {\bf q}_1'^2 }
\right)^{\frac{\bar \alpha}{2}\left(1-\frac{\zeta_2}{2} {\bar \alpha}\right) {\rm Y}} e^{\frac{3}{4} \zeta_3 {\bar \alpha}^2 {\rm Y}}
\Bigg\{\delta^{(2)} \left({\bf q}_1-{\bf q}_2\right) \nonumber\\
&&\hspace{-3.5cm}+\sum_{n=1}^\infty \prod_{i=1}^n \,  \Bigg[ \int d^2 {\bf k}_i 
{{\bar \alpha} \over 4} \left(1-\frac{\zeta_2}{2} {\bar \alpha}\right) 
 {\theta \left({\bf k}_i^2 - \lambda^2\right) \over \pi {\bf k}_i^2}
\Bigg(1+{\left({\bf q}_1'+\sum_{l=1}^{i-1}{\bf k}_l\right)^2 ({\bf q}_1+\sum_{l=1}^{i}{\bf k}_l)^2 - {\bf q}^2 {\bf k}_i^2 \over ({\bf q}_1'+\sum_{l=1}^{i}{\bf k}_l)^2 \left({\bf q}_1+\sum_{l=1}^{i-1}{\bf k}_l\right)^2}\Bigg) \nonumber\\
&&+ \Phi \left({\bf q}_1+\sum_{l=1}^{i-1}{\bf k}_l,{\bf q}_1+\sum_{l=1}^{i}{\bf k}_l\right)\Bigg]
\delta^{(2)} \left({\bf q}_1+\sum_{l=1}^n {\bf k}_l-{\bf q}_2\right)  \nonumber\\
&&\hspace{-3cm} \times\int_0^{y_{i-1}} \hspace{-.3cm} d y_i 
\left(\left({\bf q}_1+\sum_{l=1}^{i-1}{\bf k}_l\right)^2 \over 
\left({\bf q}_1+\sum_{l=1}^{i}{\bf k}_l\right)^2 \right)^{1+ {{\bar \alpha} y_i \over 2}\left(1-\frac{\zeta_2}{2} {\bar \alpha}\right) }
\left(\left({\bf q}_1'+\sum_{l=1}^{i-1}{\bf k}_l\right)^2 \over 
\left({\bf q}_1'+\sum_{l=1}^{i}{\bf k}_l\right)^2 \right)^{{ {\bar \alpha} y_i \over 2}\left(1-\frac{\zeta_2}{2} {\bar \alpha}\right) }\Bigg\}, 
\end{eqnarray}
where $y_0 \equiv Y$.
We have checked that the functions ${\cal H}$ and ${\cal F}$ are $\lambda$ independent for small 
$\lambda$.

An interesting question is to study the convergence of the sum defining the 
function ${\cal H}$ in Eq.~(\ref{H}). For a fixed value of Y and the coupling 
${\bar \alpha}_s$ a 
finite number of terms in the sum is needed to reach a good accuracy for the 
gluon Green's function. 
\begin{figure}[htbp]
\hspace{-.7cm} \includegraphics[width=8cm,angle=0]{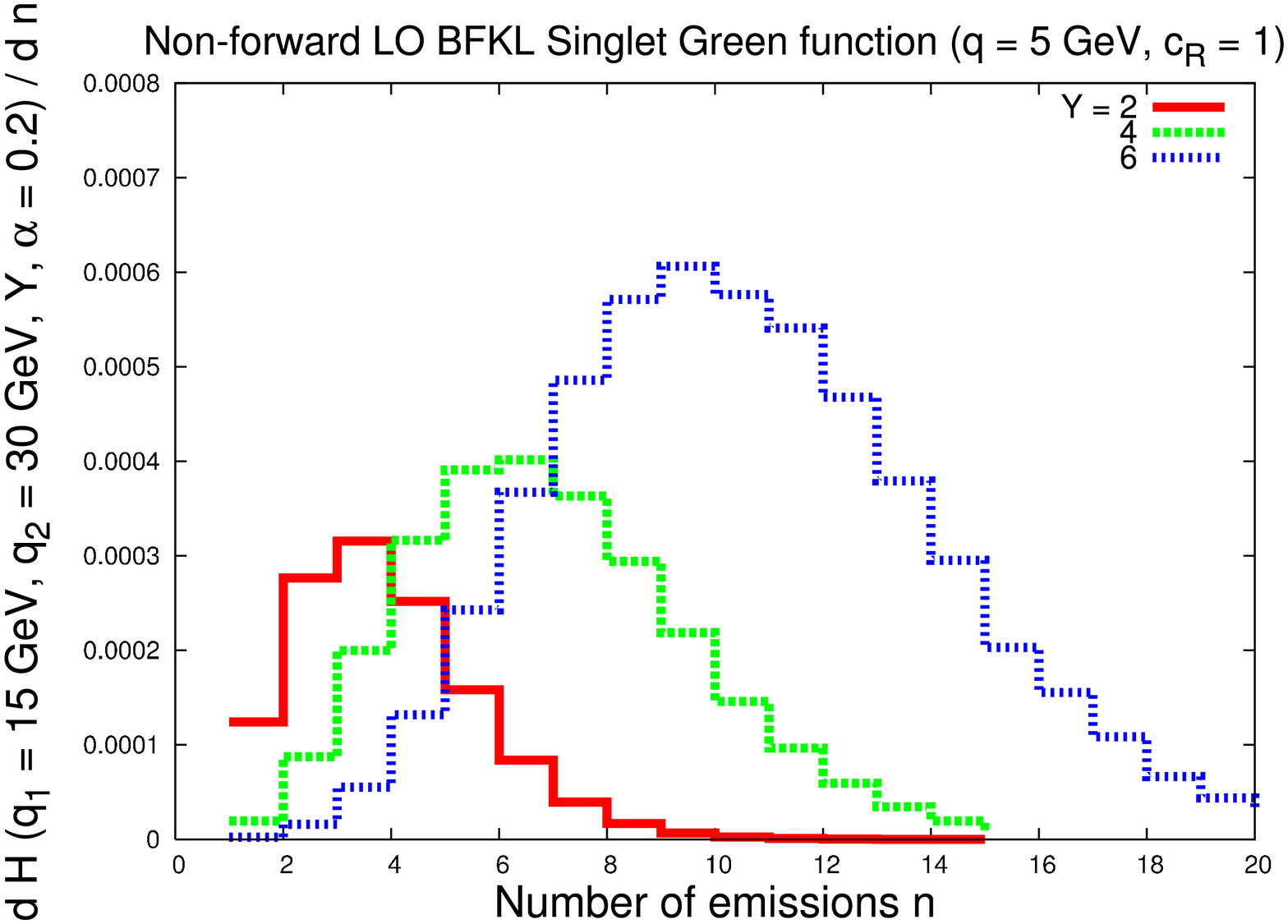}
\includegraphics[width=8cm,angle=0]{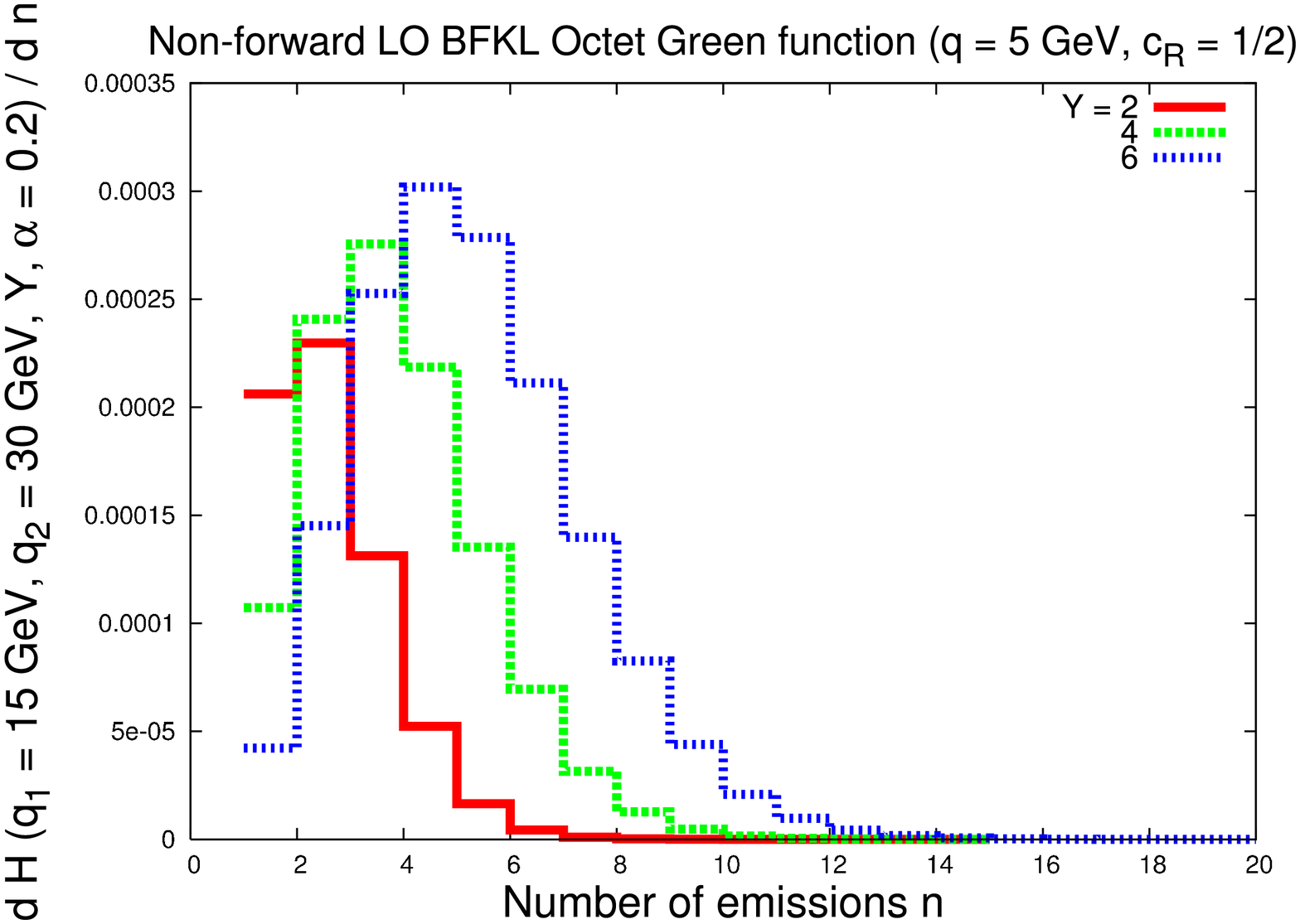}
 \caption{Distribution in the contributions to the LO BFKL gluon Green function 
    with a fixed number of iterations of the kernel, plotted for different values of the 
    center-of-mass energy, and a fixed ${\bar \alpha}_s = 0.2$.}
  \label{NumberOfTerms}
\end{figure}
As the value of the effective parameter 
${\bar \alpha}_s {\rm Y}$ 
gets larger the Green's function is more sensitive to high multiplicity terms, following 
a Poissonian distribution as it can be seen in Fig.~\ref{NumberOfTerms}.  
It is very important that 
the convergence is much better in the octet case where it is possible to get the Green's 
function with a small number of terms, see Fig.~\ref{NumberOfTerms} (right).  The same behavior 
is also observed at NLO level.
This can be qualitatively understood if we 
think of the ``real emission'' terms as pushing the Green's function to grow with $Y$. 
In the case of the color singlet state
they carry the same coefficient as the contributions from the gluon 
Regge trajectories, which make the Green's 
function  decrease with $Y$. For the adjoint representation this 
balance is broken in favor 
of the ``virtual contributions'', 
making the Green's function to grow much slower with an effective reduction in the number of emissions.

\section{Conclusions and Outlook}

As was mentioned in the introduction, our long-term perspective is to set up
a  study project focused on the LHC phenomenology of 
the BKP Odderon. We want to
study multijet events at the LHC with a Monte Carlo tool that keeps as much
exclusive information as possible. For that, the first step is to solve the BKP
equation in an iterative fashion similar to the one described here for the
solution of the BFKL equation in the adjoint representation.

The BFKL dynamics can be described, in a pictorial form,  by
a system of two reggeized gluons exchanged in
the $t$-channel which interact with each other through ordinary gluons.
This is the usual view in terms of ladder-type diagrams in which the
reggeized gluons play the role of the rails of the ladder and the
exchanged ordinary gluons are the rungs. 
If an exchange of an ordinary gluon takes place at rapidity $y_i$ then the next one
is allowed to happen at rapidity $y_{i+1}$ further down the ladder.
That simply means that in order to use an iterative solution, one has to increase
the number of rungs by one in each iteration until convergence is reached. 

In the BKP framework, this picture has to be modified. 
There are now three reggeized gluons
exchanged in the $t$-channel and they can interact, locally in rapidity,
in pairs through
the exchange of ordinary gluons. The ladder is no more one with two rails but rather one with
three rails and each rung can connect any two of them.
The whole system is in the color singlet representation whereas any
subset of two reggeized gluons is in the symmetric color octet representation. 
There would not be hardly any hope to  pursue an iterative solution
through the Monte Carlo approach if it were not for the fact that any pair
of the three reggeized gluons is in the color octet representation which,
as we saw in the previous section, leads to a much faster convergence
compared to the color singlet state.

This is a key feature in our approach for finding an
iterative solution of the BKP equation.
We will present our results with regard to that elsewhere~\cite{BKP}.

\section*{Acknowledgments}
We thank Victor Fadin and Lev Lipatov for discussions. G. C. thanks the Department of Theoretical Physics at the Aut{\'o}noma University of Madrid and the ``Instituto de F{\'\i}sica Te{\' o}rica 
UAM / CSIC'' for their hospitality. 
We thank the CERN PH-TH Unit where the first stages of setting up the BKP project took place. 
We acknowledge support by the
Research Executive Agency (REA) of the European Union under 
the Grant Agreement number PITN-GA-2010-264564 (LHCPhenoNet),
the Comunidad de Madrid
through Proyecto HEPHACOS ESP-1473, and MICINN (FPA2010-17747),
the Spanish Government and EU ERDF funds 
(grants FPA2007-60323, FPA2011-23778 and CSD2007-00042 
Consolider Project CPAN) and by GV (PROMETEUII/2013/007). 
G.C. acknowledges support from Marie Curie actions (PIEF-GA-2011-298582).

\end{document}